\def\year{2019}\relax
\documentclass[letterpaper]{article} 
\usepackage{aaai19}  
\usepackage{times}  
\usepackage{helvet}  
\usepackage{courier}  
\usepackage{url}  
\usepackage{graphicx}  
\usepackage[draft=false]{hyperref}
\graphicspath{}
\usepackage[font=small,labelfont=bf]{caption}

\begin{document}
%
\title{Towards Task Understanding in Visual Settings}
\author{Sebastin Santy, Wazeer Zulfikar\\
BITS Pilani KK Birla Goa Campus, India\\
f20150\{357, 003\}@goa.bits-pilani.ac.in\\
+1(412)708-5517, +91-7774031034
\And Rishabh Mehrotra\\
Spotify Research\\
London, United Kingdom\\
rishabhm@spotify.com
\And
Emine Yilmaz\\
University College London\\
London, United Kingdom\\
emine.yilmaz@ucl.ac.uk
}

\maketitle
\begin{abstract}
We consider the problem of understanding real world tasks depicted in visual images. While most existing image captioning methods excel in producing natural language descriptions of visual scenes involving human tasks, there is often the need for an understanding of the exact task being undertaken rather than a literal description of the scene. We leverage insights from real world task understanding systems, and propose a framework composed of convolutional neural networks, and an external hierarchical task ontology to produce task descriptions from input images. Detailed experiments highlight the efficacy of the extracted descriptions, which could potentially find their way in many applications, including image alt text generation.

\end{abstract}

\begin{figure*}[h!]
    \centering
    \begin{minipage}{0.33\textwidth}
        \centering
        \includegraphics[width=\textwidth, draft=false]{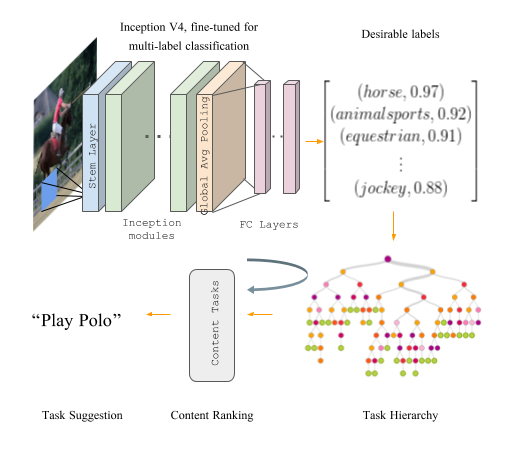} 
        \vspace{-3mm}
        \caption{Architecture}
        \label{fig:arch}
        \vspace{-5mm}
    \end{minipage}\hfill
    \begin{minipage}{0.33\textwidth}
        \centering
        \vspace{8mm}
        \includegraphics[width=\textwidth, draft=false]{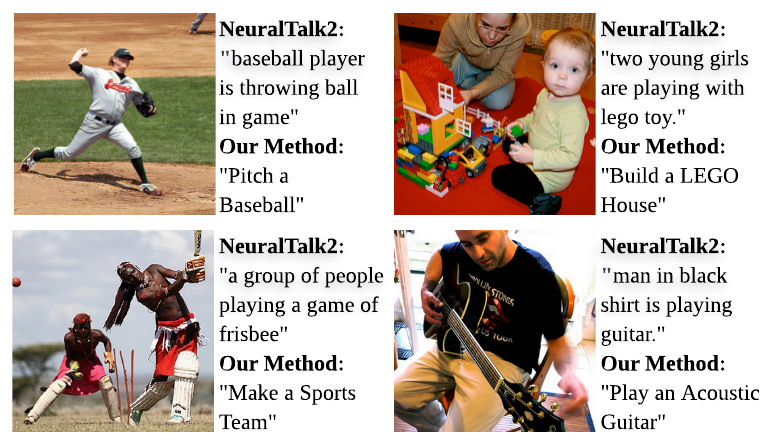} 
        \vspace{3mm}
        \caption{Description Comparisons}
        \label{fig:taskdesc}
    \end{minipage}
    \begin{minipage}{0.33\textwidth}
        \centering
        \includegraphics[width=\textwidth, draft=false]{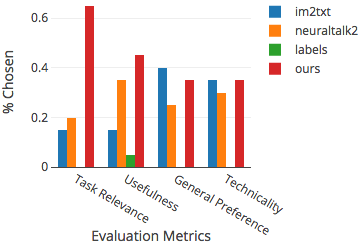} 
        \vspace{-3mm}
        \caption{Crowdsourced evaluation compared}
        \label{fig:eval}
        \vspace{-8mm}
    \end{minipage}
\end{figure*}
\vspace{-6mm}
\section{Introduction}
A substantial portion of real world images depict a human task; for example, Figure \ref{fig:taskdesc} shows tasks like pitching a baseball, building lego toys or playing a guitar. While humans are efficient at understanding and describing the task intent by just a quick glance at a visual scene, most image captioning systems are only able to generate plain description of the different visual elements in the image. Especially in the cases of highly complex tasks, predicting task intent may involve more than just generating a plain description of the scene. Understanding the task context in visual scenes is indeed important in a number of application settings, including image alt text generation, image suggestions or image search. 


While a lot of work has gone into generating image descriptions, most prior work 
\cite{bernardi2016automatic} have used visual and multi-modal space to assist the generation of a dense natural language description which allow for a more expressive prediction. Often times such detailed descriptions of the different visual elements are not required but a minimal explanation as to what is happening in the image suffices.
Such \textit{task}-based captions help in keeping the description more technical and contextual. For example, ``Pitch a Baseball" is an apt minimal replacement to ``baseball player is throwing ball in game." in Figure \ref{fig:taskdesc}, while being more technical, contextual and maintaining brevity. Our method primarily aims to improve existing methods by specifically overcoming this limitation by trying to precisely predict the task present in the scene while simultaneously preserving the context, as opposed to synthesizing a verbose description.

We jointly leverage insights from recent advancements in deep convolutional architectures and hierarchical task ontologies, and propose a two phase model to suggest scene task descriptions. The convolutional architecture generates contextual labels from the image, while the task extractor maps these labels to real world tasks. We leverage the TaskHierarchy138K \footnote{https://usercontext.github.io/TaskHierarchy138K/} ontology which contains `\textit{tasks}' and keywords associated with each of these `\textit{tasks}' in a hierarchical structure, with a complex task often decomposed into simpler sub-tasks.
Detailed experiments based on both qualitative and quantitative experiments demonstrate that our method not only helps in extracting task information, but also provides more useful descriptions when compared with state-of-the-art image description approaches.


\section{Approach}
In order to extract the tasks depicted in an image, we propose a two phased model: i) Multi-label classification of scenes to generate input labels for the task extractor  and ii) Leveraging external hierarchical ontology for task identification by task extractor.

For image classification, we train the deep Inception v4 architecture \cite{szegedy2017inception} which is capable of detecting 1000 categories and fine-tune the network for multi-label contextual classification of the scene. The input image is fed to this classifier to obtain contextual labels along with their respective confidence scores. The labels generated by the classifier is further processed to filter out redundant information, and the resulting filtered set of labels is then passed to the task extractor module.
Given the labels of the image, the task extractor will probe into a task hierarchy to suggest tasks . In order to infer the task, we leverage TaskHierarchy138K, which is an external task ontology that uses Wikihow articles to provide task information for over 100k real world tasks. Each node in this hierarchy represents a WikiHow category, with its children nodes representing its subcategories. A node $k$ contains a Representative embedding ${h_{R}}_{k}$ and an Average Embedding ${h_{A}}_{k}$. ${h_{R}}_{k}$ is an average embedding of the articles present in node $k$. ${h_{A}}_{k}$ is $\frac{1}{n}\sum_{i=1}^{n = |children(k)|}{h_{A}}_{i}$ which is recursively calculated for all nodes except leaf-nodes. For leaf-nodes, ${h_{R}}_{k} = {h_{A}}_{k}$.

Given a task hierarchy and a set of labels produced from the classifier, we start with the root of the tree, and then trickle the labels down through the hierarchy. This trickling process is divided into two steps in order to achieve speed while simultaneously making it robust to noise. 
\begin{enumerate}
    \item \textbf{First Order Trickling}: We pass down a weighted average vector embedding ($h_{l}$) of all the labels starting from node $k$ by recursively trickling to the child node with maximum cosine similarity ($cosine$) between $h_{l}$ and ${h_{A}} \in children(k)$. The trickling stops at node $k'$ when the $cosine(h_{l}, {h_{R}}_{k'})\geq \forall cosine(h_{l}, {h_{A}}\in children(k'))$ where the $cosine(h_{l}, {h_{R}}_{k'})$ acts as a threshold.
    \item \textbf{Second Order Trickling}: This is used for further trickling down the achieved node in the previous step. Some specific low-weighted labels belonging to a subcategory of the achieved node gets buried in $h_{l}$. Hence, we calculate $cosine$ between each label $l$ and $\forall {h_{A}} \in children(k)$. The labels are trickled down to the node which returns the maximum $cosine$, iff it is higher than threshold defined in the previous step.
\end{enumerate}
After the node is captured by the trickling process, we rank the content tasks using cosine similarity over their respective article knowledge, to suggest the appropriate task.

\section{Results and Discussion}
To the best of our knowledge, this is the first work done on predicting tasks being undertaken in a given scene. Generating expressive image descriptions is the closest work to task suggestion. We compare our results with one of the best image descriptors - NeuralTalk2\cite{karpathy2015deep} in Figure \ref{fig:taskdesc}. It should be noted that our work does not compete with the image descriptor. We want to make the reader aware that we are able to suggest task fairly accurately with a less complex model by leveraging existing task information.

We conducted a crowd-sourced study on Amazon Mechanical Turk. In the study, workers answer 10-randomly picked images along with image descriptions generated by NeuralTalk2, im2txt\cite{vinyals2015show}, multi-label classifier \cite{szegedy2017inception} (as baselines) and our method. NeuralTalk2 uses convolutional and recurrent neural networks in multimodal space to generate image descriptions. im2txt is similar to NeuralTalk2 with a better classifier. We evaluate on the basis of 4 metrics: Task Relevance, Usefulness, General Preference and Technicality. As seen in Figure\ref{fig:eval}, our method outweighs NeuralTalk2 and im2txt captions for task relevance metric by a large margin and performs almost equally for the other three metrics. As expected, multi-label classifier tags perform poorly due to non-aesthetic descriptions. This study reinforces our assertion on task suggestion capability of our method. 


\section{Conclusion and Future Work}
In this work, we propose a novel method for a scene task suggestion system. These descriptions can be used for applications like image alt text generation or as priors to existing image description models to build their descriptions upon, rather than generating them base up. However, this kind of a system is constrained to work on scenes where the task being done is a prominent part of it. We intend to extend this work to aid in the existing dense image description generation, making models intrinsically more task-aware by injecting task coherence scores within their architecture.

\section{Acknowledgements}
This project was partially funded by the EPSRC Fellowship titled ``Task Based Information Retrieval", grant reference number EP/P024289/1.

\bibliographystyle{aaai}
\bibliography{bibliography}
\end{document}